# Understanding College Students' Phone Call Behaviors Towards a Sustainable Mobile Health and Wellbeing Solution

Yugyeong Kim[1], Sudip Vhaduri[1], and Christian Poellabauer[2]
[1]{ykim180, svhaduri}@fordham.edu, [2]cpoellab@nd.edu
[1]Fordham University, NY, USA  [2]University of Notre Dame, IN, USA

## Abstract

During the transition from high school to on-campus college life, a student leaves home and starts facing enormous life changes, including meeting new people, more responsibilities, being away from family, and academic challenges. These recent changes lead to an elevation of stress and anxiety, affecting a student's health and wellbeing. With the help of smartphones and their rich collection of sensors, we can continuously monitor various factors that affect students' behavioral patterns, such as communication behaviors associated with their health, wellbeing, and academic success. In this work, we try to assess college students' communication patterns (in terms of phone call duration and frequency) that vary across various geographical contexts (e.g., dormitories, classes, dining) during different times (e.g., epochs of a day, days of a week) using visualization techniques. Findings from this work will help foster the design and delivery of smartphone-based health interventions; thereby, help the students adapt to the changes in life.

## 1. Introduction

Due to the recent advancement of technology, smartphones have become an indispensable part of our life. With the ever increasing sensing capabilities of smartphones, we can track different spatio-temporal factors and their effect on our various behavioral patterns, such as physical activity and communication patterns, which can be used to develop predictive models and deliver interventions (Amin *et al.*, 2009; Vhaduri *et al.*, 2019). Similarly, with the help of smartphones' sensing capability, we can monitor college students' daily life to find factors to improve their lives. Students leave their school friends and homes for the first time when starting on-campus college life. During this transition period, students go through a stage where they may experience extreme stress, anxieties, and depression (Montoliu *et al.*, 2013; Bogomolov *et al.*, 2014; Vhaduri *et al.*, 2018d) as they experience many changes, such as adapting to a new lifestyle, meeting new people, and facing academic deadlines. These changes impact their mental and physical health and academic performance (Trockel *et al.*, 2000). Therefore, a better understanding of various temporal and geographical patterns of college students' communication behaviors can help us to design and deliver different smart health interventions to ensure students' health and wellbeing while adapting to a new lifestyle during the transition period.

In this paper, we present our analysis of college students' communication patterns (in terms of call duration and frequency/count) during various temporal contexts (e.g., time of a day, day of a week) in different geographical contexts using a dataset collected from a cohort of more than 400 on-campus first-year students over three consecutive semesters. This work's primary goal is to present the insights obtained from the visualization of communication patterns that vary across different contexts. This work provides a foundation for potential future strategies that aim to improve the physical and mental wellbeing and academic success of students who leave their home after high school to start their on-campus undergraduate studies. Furthermore, this can be utilized to design, develop, and deploy smartphone-based health interventions to help students to cope with changes and challenges that students face while transitioning between school and college.

## 3. Dataset

The *NetHealth* mobile crowdsensing study (Vhaduri *et al.*, 2016, 2018c, 2017c, 2017d) began at the University of Notre Dame in 2015 with over 400 on-campus freshmen (average age of 17 years and 11 months with a standard deviation of 11 months) to investigate the impacts of "always-on connectivity" on the health habits, emotional wellness, and social ties of college students over multiple semesters. All procedures were fully approved by the IRB before distribution. Students were recruited using both e-mail and Facebook invites and were given a data collection app for their iPhones (Hossain *et al.*, 2016). The smartphone app collected data 24 hours a day. The data collected by the smartphone app includes identifiers of the device's network connections (Wi-Fi, cellular), device state (e.g., battery charge level), screen state, geographic location, and user communications (e.g., phone calls). While some sensor data, such as Wi-Fi and cellular, are recorded at a fixed sampling frequency, some data, such as the phone call data (i.e., start time and duration), are recorded using a callback mechanism. Location data is recorded with a sampling period of 165 seconds, and it comes as a series of location points. Each location point is defined as a tuple $\{\lambda, \vartheta, T\}$ (Montoliu *et al.*, 2013), where $\lambda$ and $\vartheta$ are the latitude and longitude of a location point, and $T$ is the timestamp when the location point is recorded. While most sensor data are transmitted to a remote server during a nightly upload, some data, such as communication data (phone calls, texts, etc.) and statistics, are transmitted via a desktop client. In this work, we use location data and phone call data that come with the time of call start, duration (measured in seconds), flags to check whether the call is answered and originated/received, and phone number on the other end.

## 2. Methodology

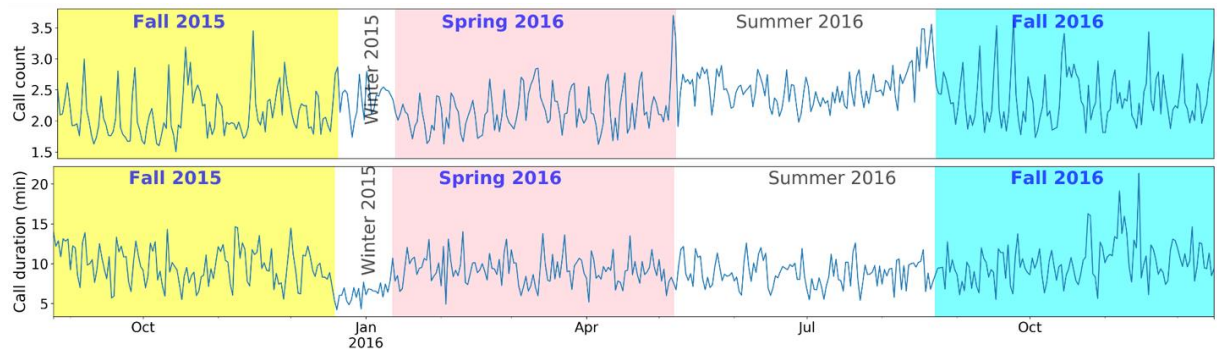

Figure 1: Time series of day-level average phone call count (top) and duration (bottom)

In this paper, we intend to demonstrate the link between phone call behaviors in various temporal and geographical contexts of a cohort of on-campus first-year students. Before we describe the detailed analysis, we first introduce phone call related terminologies, and then we introduce the dataset and methods used in this work.

We primarily use two communication measures, namely "call count" and "call duration" (Figure 1), to assess students' communication behaviors. They are defined as below:

- Average call duration, i.e., call duration per person in a unit time (e.g., in a day), is computed as:

$$D^j = \frac{\sum_{i=1}^{n^j} d_i^j}{n^j}$$

Where $d_i^j$ is the call duration of $i^{th}$ person in the $j^{th}$ day, $n^j$ is the number of active persons in the $j^{th}$ day, and $D^j$ is the call duration per person in the $j^{th}$ day. Finally, the call duration per person per day is the average per person call duration obtained from all days.

- Average call count, i.e., call count per person in a unit time (e.g., in a day), is computed as:

$$C^j = \frac{\sum_{i=1}^{n^j} c_i^j}{n^j}$$

Where $c_i^j$ is the call count of $i^{th}$ person in the $j^{th}$ day, $n^j$ is the number of active persons in the $j^{th}$ day, and $C^j$ is the call count per person in the $j^{th}$ day. Finally, the call count per person per day is the average per person call count obtained from all days.

Additionally, we use different terminologies defined as below:

- Places of interest (POIs) are the places where people either spend a significant amount of their day or frequently visit (Vhaduri *et al.* 2017a). For example, for an on-campus freshmen cohort, POIs can be dormitories (DM), class buildings (CL), dining halls and food courts (DI), athletic facilities (AL), as well as other indoor (OI) and outdoor places (OD) (Vhaduri *et al.*, 2018a, 2018b).

- To better understand temporal patterns of communication behaviors, we split the entire day into four major epochs: morning (8 am – 12 pm or 8 – 12 in 24-hour clock), afternoon (12 pm – 6 pm or 12 – 18 in 24-hour clock), evening (6 pm – 12 am or 18 – 24 in 24-hour clock), and night (0 – 7 am or 0 – 7 in 24-hour clock) epochs (Vhaduri *et al.*, 2016, 2017a).

## 4. Results

This section presents geo-temporal analysis using bubble charts, where bubble placement in the y-axis shows the average call duration (in minutes) or count and size presents the standard deviation of call duration or the count. In Figure 2, we observe that the dormitories are the dominant POIs where most students make more prolonged and frequent phone calls across the four epochs in a day followed by outdoors. The result is very intuitive since students spend a significant part of their day in dormitories.

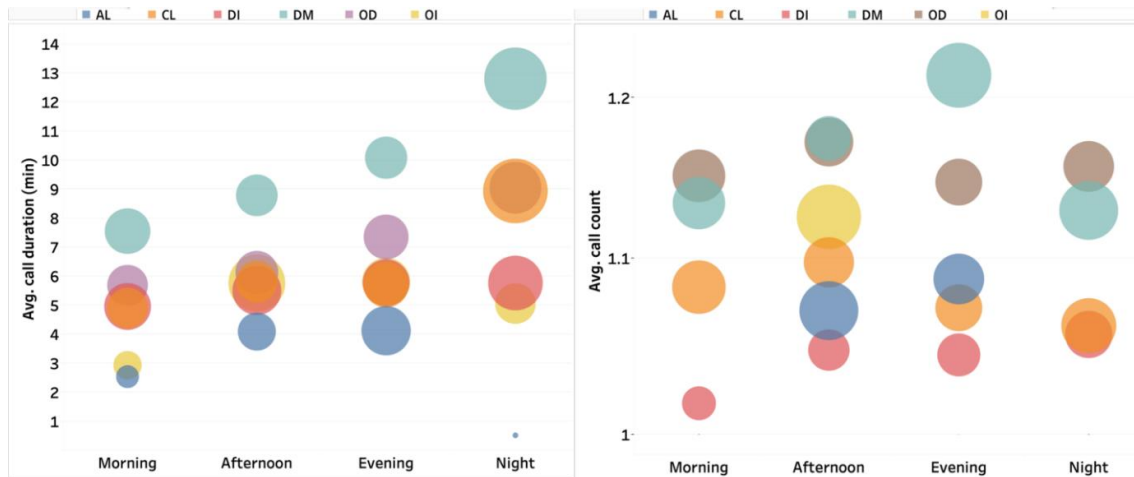

Figure 2: Bubble charts of average (a) call duration (min) (left) and (b) call count (right) across various on-campus POIs during different epochs of a day

In the evening, the second-highest call duration comes from outdoor POIs (ODs). We observe that during the night, the average call duration becomes maximum. The outdoor POIs (ODs) are always the second dominant group in terms of average call duration and count, and we observe a steady pattern across all epochs. In Figure 2, we observe that evening epochs at dormitories (DM) achieve the highest average call count. In the figure, we further observe that students make shorter and more frequent calls in the evenings while making longer and less regular calls at nights, particularly in DMs. Additionally, the standard deviation is higher at night, which shows that call duration varies significantly across individuals during this epoch.

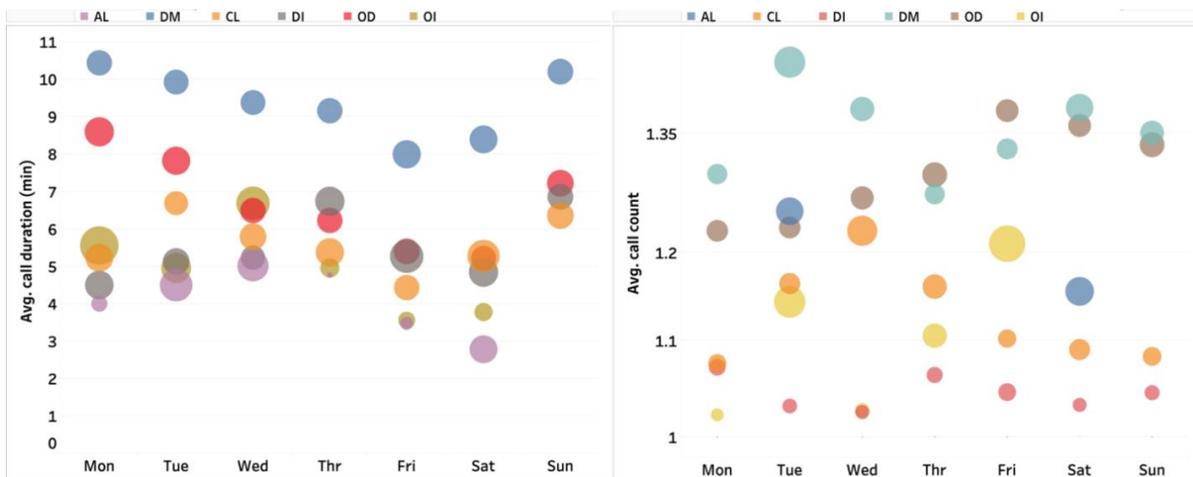

Figure 3: Bubble charts of average (a) call duration (min) (left) and (b) call count (right) across various on-campus POIs during different days of a week.

In Figure 3, we find that dormitories are the dominating POIs for both call duration and count across the seven days of a week. In the figure, we observe that the average call duration at dormitories is significantly higher than any other POIs. Outdoor POIs (ODs) are the next dominating POI in terms of average call duration. Call duration in athletic POIs (ALs) is very low across all days of a week with very high standard deviations. Other indoor POIs (OIs) reaches the highest on Sundays because OIs include the university churches. Since primary mass is held on Sunday, a higher duration on Sunday is expected.

## 5. Conclusions

To the best of our knowledge, this is the first work that investigates the change of communication behaviors during various temporal contexts, as well in different geographical contexts using objective smartphone data collected from a cohort of more than 400 on-campus freshmen over three consecutive semesters. We find that students make longer and less frequent calls on Sundays and Mondays compared to the other days when the students make shorter and more frequent calls. This work provides a foundation for potential future strategies to develop mobile health interventions that aim at improving the physical and mental wellbeing, as well as the academic success of students who leave their home after high school to start their college studies. This work has some limitations, which we plan to address in the future. In the future, we plan to perform a more detailed analysis of college students' communication patterns across different geographical and temporal contexts using detailed visualization and statistical tests.